\documentclass[useAMS,usenatbib]{mn2e}

\usepackage{graphicx}
                                                 
\newcommand{\mum}{$\mu$m}

\title[Heating of  NEOs and meteoroids]{Heating  of near-Earth objects
  and meteoroids  due to close approaches to the Sun} 
\author[]{S. Marchi$^{1}$\thanks{E-mail:simone.marchi@unipd.it},  
M. Delbo'$^{2}$, 
A. Morbidelli$^{2}$, 
P. Paolicchi$^{3}$,
M. Lazzarin$^{1}$\\
$^{1}$ Dipartimento di Astronomia, Universit\`a di Padova, I-35122  Padova,   Italy\\
$^{2}$ Observatoire de la Cote d'Azur, F-06304  Nice,   France\\
$^{3}$ Dipartimento di Fisica, Universit\`a di Pisa, I-56127 Pisa, Italy}

\begin{document}

\date{Submitted 19 June 2009, Accepted 28 July 2009}


\maketitle


\begin{abstract}

It  is  known that  near-Earth  objects  (NEOs)  during their  orbital
evolution may often undergo close  approaches to the Sun. Indeed it is
estimated that  up to $\sim$70\%  of them end their  orbital evolution
colliding with the Sun.  Starting from the present orbital properties,
it is  possible to  compute the most  likely past evolution  for every
NEO, and  to trace its  distance from the  Sun.  We find that  a large
fraction of the  population may have experienced in  the past frequent
close  approaches,   and  thus,  as  a   consequence,  a  considerable
Sun--driven heating,  not trivially correlated to  the present orbits.
The  detailed  dynamical  behavior,  the rotational  and  the  thermal
properties of NEOs determine the exact amount of the resulting heating
due to the Sun.\\
In the present  paper we discuss the general  features of the process,
providing estimates of the  surface temperature reached by NEOs during
their evolution.  Moreover, we investigate the effects of this process
on  meteor-size bodies,  analyzing possible  differences with  the NEO
population.   We also  discuss some  possible effects  of  the heating
which can be observed  through remote sensing by ground--based surveys
or space missions.

\end{abstract}


\section{Introduction}

It is well  known that the physical properties  of most asteroids (and
their  fragments,  the  meteorites)  have  been  affected  by  heating
processes.  The analysis of  the available meteorite collections shows
that meteorites and their asteroidal parent bodies have been thermally
altered to  some degree.  The  alteration ranges from  low temperature
aqueous  processes  ($<$300~K)  up  to partial  or  complete  melting,
differentiation    and    fractional    crystallization    ($>$1150~K)
\citep{kei00}.   The main  source of  heating has  been  identified as
internal  heating  induced  by  decay  of  short-lived  radionuclides,
notably $^{26}$Al \citep{gri93}. The effects of heating processes have
also been  identified on the  surfaces of asteroids by  remote sensing
spectroscopy of primitive main belt asteroids \citep{hir93, vil89}.
Despite  the   general  consensus  on   radionuclides  heating,  other
processes have  been suggested to  occur: for instance  heating during
the preaccretion  phase or due  to collisions. The latter  may produce
high  temperatures in  the regions  involved in  high  energy impacts,
often causing  the partial melting  of the involved  bodies \citep[see
  discussion in][]{riv02}.\\
In  this  paper we  discuss  a new  heating  process,  which has  been
neglected so  far, namely the heating  due to close  approaches to the
Sun.   This effect  may be  relevant for  bodies such  as  NEOs, whose
perihelia  are  sometimes  very  close  to the  Sun,  and  significant
temperatures can be attained. \\
In the present  NEO population the fraction of  bodies with relatively
small perihelion  ($q$) is very small.   However, this is  by no means
representative of  the past  (and future) dynamical  tracks.  Detailed
simulations  show that a  large fraction  of NEOs  may have  had small
perihelion  distances for  some time,  hence experiencing  episodes of
strong heating.  Moreover, since  meteoroids have an orbital evolution
similar to  that of NEOs, this  process may also  affect the meteorite
samples.\\  
The detailed effects of the heating during Sun close approaches depend
on  the  thermal  properties   of  the  bodies.   Albedo,  emissivity,
macroscopic  surface roughness  and thermal  inertia are  the physical
parameters  that   determine  the  surface   temperature.  The  latter
parameter  measures  the  resistance  of  a  material  to  temperature
changes, e.g.  due to the varying day/night illumination caused by the
rotation of the body.  The rotation state of the body also affects its
temperature profile.  For instance,  for given set  of thermo-physical
parameters, a  rapid rotating asteroid has  a temperature distribution
more  smoothed out  in  longitude  than a  slower  rotating one.   The
direction of the body's spin vector causes also seasonal effects.\\
In this  work we estimate  the amount of  heating due to the  Sun that
NEOs  and meteorites are  likely to  have experienced.  We do  this by
studying  their  dynamical history.  We  then  use  thermal models  to
correlate their dynamical  history with surface temperatures. Finally,
some possible consequences of the heating are discussed.

\section{Orbital properties of NEOs}

NEOs  evolve   rapidly  in  the  Keplerian   orbital  elements  space,
$(a,e,i)$,  due  to  a  combination  of  close  encounters  with  the
terrestrial planets and resonances with the giant planets. Two aspects
of this orbital evolution are relevant for the analysis of the heating
process:  (i) The probability  that each  NEO attained  a $q$  below a
given  limit ($q_s$) during  the evolution;  (ii) The  cumulative time
spent at distances  below $q_s$.  In the following  sections, we first
explain how we estimated these quantities and then present and discuss
the results that we have obtained.

\subsection{Estimating the past dynamical histories of NEOs}

The most  complete model of the  orbital distribution of  NEOs is that
developed in  \citet{bot00, bot02}. This model assumes  that NEOs come
primarily from  5 intermediate sources: the  3:1 mean-motion resonance
with  Jupiter,  the   $\nu_6$  secular  resonance,  the  Mars-crossing
population  (IMC), the  collection  of resonances  crossing the  outer
asteroid belt (OB) and  the trans-Neptunian region (as dormant Jupiter
family  comets; JFC).   In  the NEO  orbital  space ($q<1.3$~AU),  the
steady state  orbital distribution of  the bodies coming from  each of
these sources  ($R_{\rm source} (a,e,i)$)  was computed using  a large
number of numerical simulations.   The orbital distribution of the NEO
population  was  then  constructed  as  a linear  combination  of  the
distributions related to each source, that is:
$$ R_{\rm NEO} (a,e,i)=\sum_{\rm source} N_{\rm
 source} R_{\rm source} (a,e,i)\ . $$  
The coefficients $N_{\rm source}$ for this combination were determined
by  fitting the distribution  of the  NEOs discovered  or accidentally
recovered by the Spacewatch survey, once observational biases had been
taken  into  account.   This   model  was  later  shown  to  reproduce
adequately also  the orbital distributions  of NEOs discovered  by the
LINEAR and Catalina Sky  surveys \citep{zav08}.

Given a  specific NEO, using the  Bottke et al. model  we estimate the
probability  that it comes  from each  of the  considered intermediate
sources, as follows. We first  define a box in $(a,e,i)$ orbital space
in which the NEO currently  resides. Then, for each source, we compute
$\bar R_{\rm source}$, as the integral of $R_{\rm source}(a,e,i)$ over
the box; we  also compute $\bar R_{\rm NEO}$  in an equivalent manner.
Then, the probability that the NEO  comes from a given source is given
by:
$$ P({\rm source})=N_{\rm source}\bar R_{\rm source}/\bar R_{\rm NEO}\ . $$
In  our calculation, we  use boxes  of size  0.1AUx0.1x5$^\circ$.  The
probability $P(q_s)$ that  the considered NEO achieved in  the past an
orbit $q<q_s$ is
$$ P(q_s)=\sum_{\rm  source} P({\rm source})  P_{\rm source}(q_s)\ ,$$
where $P_{\rm source}(q)$ is  the probability that objects coming from
the source  achieved $q<q_s$ {\it before} entering  the $(a,e,i)$ box.
Similarly, the mean time $T(q_s)$  spent on orbits with $q<q_s$ can be
computed as
$$ T(q_s)=\sum_{\rm source} P({\rm source}) P_{\rm source}(q_s) T_{\rm
  source}(q_s)/P(q_s)\  ,$$ 
where $T_{\rm  source}(q_s)$ is  the mean time  spent on  orbits with
  $q<q_s$  by the  particles reaching  this  kind of  orbits from  the
  source, before entering the $(a,e,i)$ box.
To  compute $P_{\rm  source}(q_s)$ and  $T_{\rm source}(q_s)$  we went
back to  the original simulations used  by Bottke et  al. to construct
the $R_{\rm source}(a,e,i)$ distributions.   The computation is just a
matter  of book-keeping, while  the outputs  of these  simulations are
read sequentially  in time.  For  simplicity (and availability  of the
original simulations)  we did  this only for  the 3:1, $\nu_6$  and OB
sources.   We  then assumed  that  the  objects  coming from  the  IMC
population behave  statistically as a  1:1 combination of  the objects
coming from the 3:1 and $\nu_6$ sources; similarly we assumed that the
objects  coming  from  the  JFC  source  share  the  same  statistical
properties of those coming from  the OB source.  These assumptions are
of  course not exact, but  are relatively  close to  reality. By
doing  this,   the  equations  defining  $P(q_s)$   and  $T(q_s)$  are
effectively restricted  to three sources  (3:1, $\nu_6$, OB)  with new
coefficients
\begin{eqnarray}
P'({\rm 3:1})&=&P({\rm 3:1})+1/2 P({\rm IMC})\ ,\cr
P'(\nu_6)&=&P(\nu_6)+1/2 P({\rm IMC})\ ,\cr
P'({\rm OB})&=&P({\rm OB})+P({\rm JFC})\ .
\end{eqnarray}

\subsection{Results}

\begin{figure*}
\includegraphics[width=17cm]{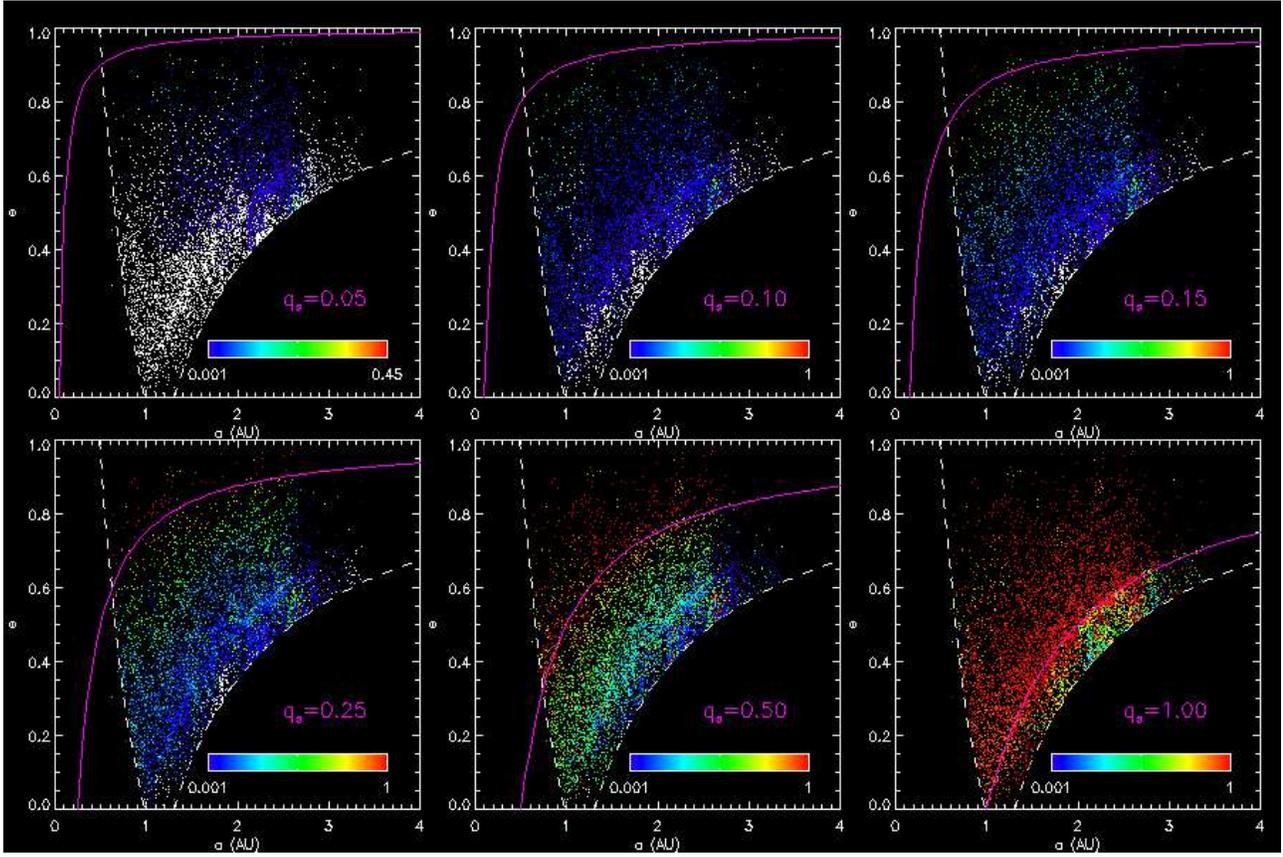}   
\vspace{0.1cm}
\caption{Semimajor axis  vs eccentricity  scatter plot of  the present
  NEO population  for several values  of $q_s$. NEOs are  plotted with
  colors coded  accordingly to their  probability of stay  below $q_s$
  (white  dots  indicate  zero  probability).   For  each  panel,  the
  corresponding $q=q_s$ curve is also shown.}
\label{a_e_prob}
\end{figure*}

\begin{figure*}
\includegraphics[width=7cm,angle=-90]{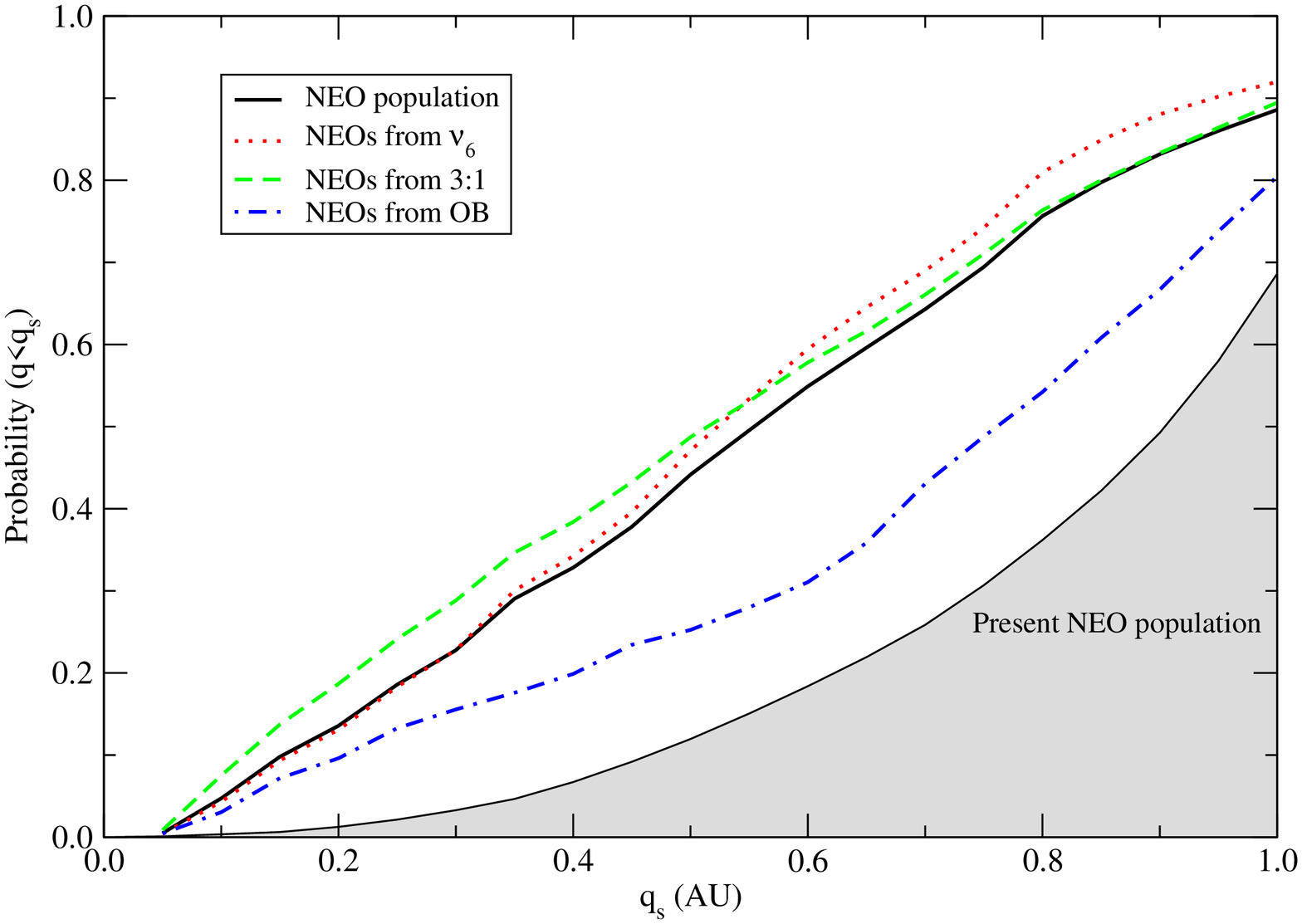}
   \includegraphics[width=7cm,angle=-90]{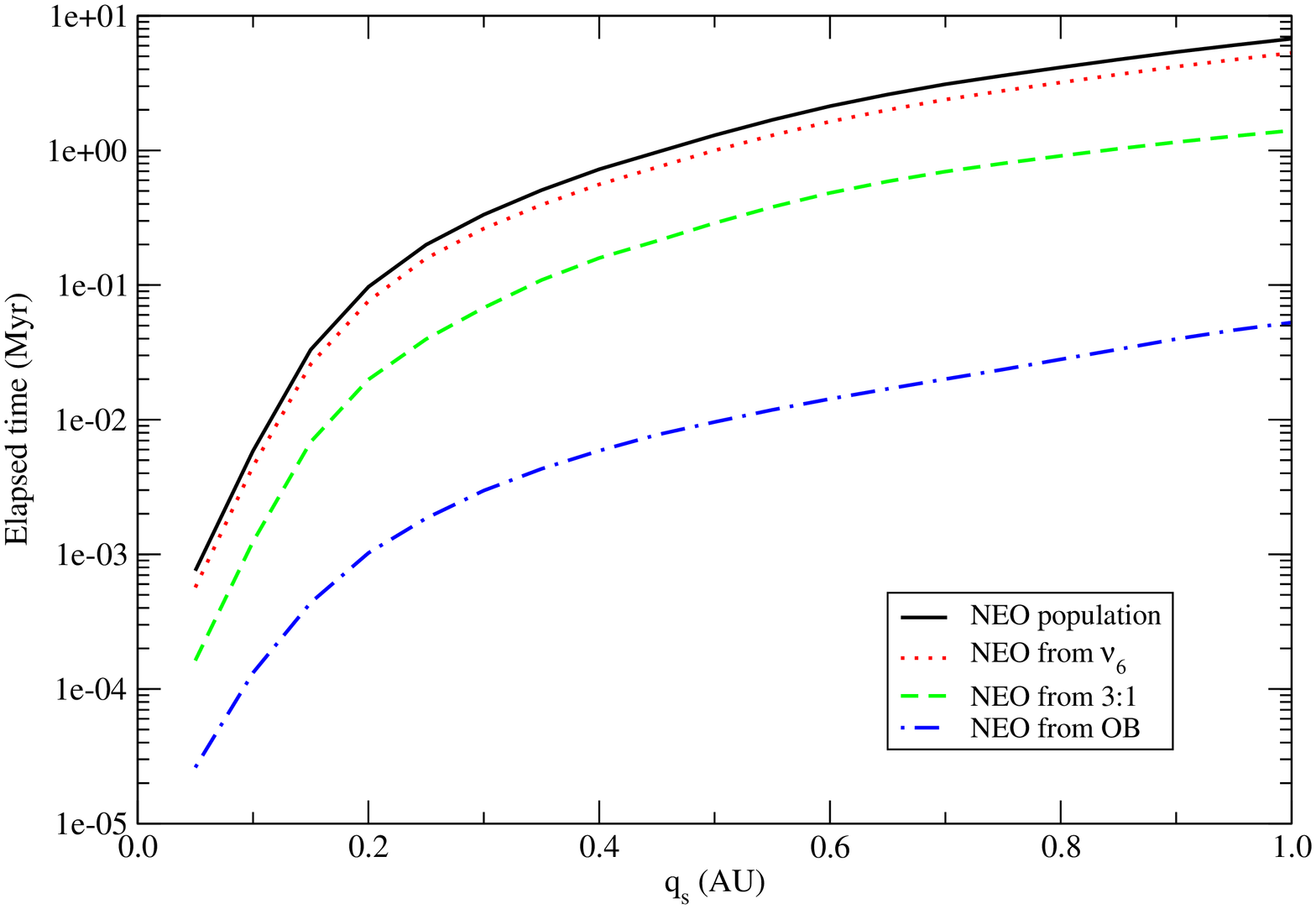}
\vspace{0.1cm}
\caption{Left panel:  average NEO probability of  $q<q_s$.  The curves
  for the three  source regions have been normalized  to the number of
  NEOs coming  from each region.  Right panel:  average NEO cumulative
  time spent at $q<q_s$.  The curves for the three main source regions
  are also shown.}
\label{prob_temp}
\end{figure*}

\begin{figure*}
\includegraphics[width=17cm]{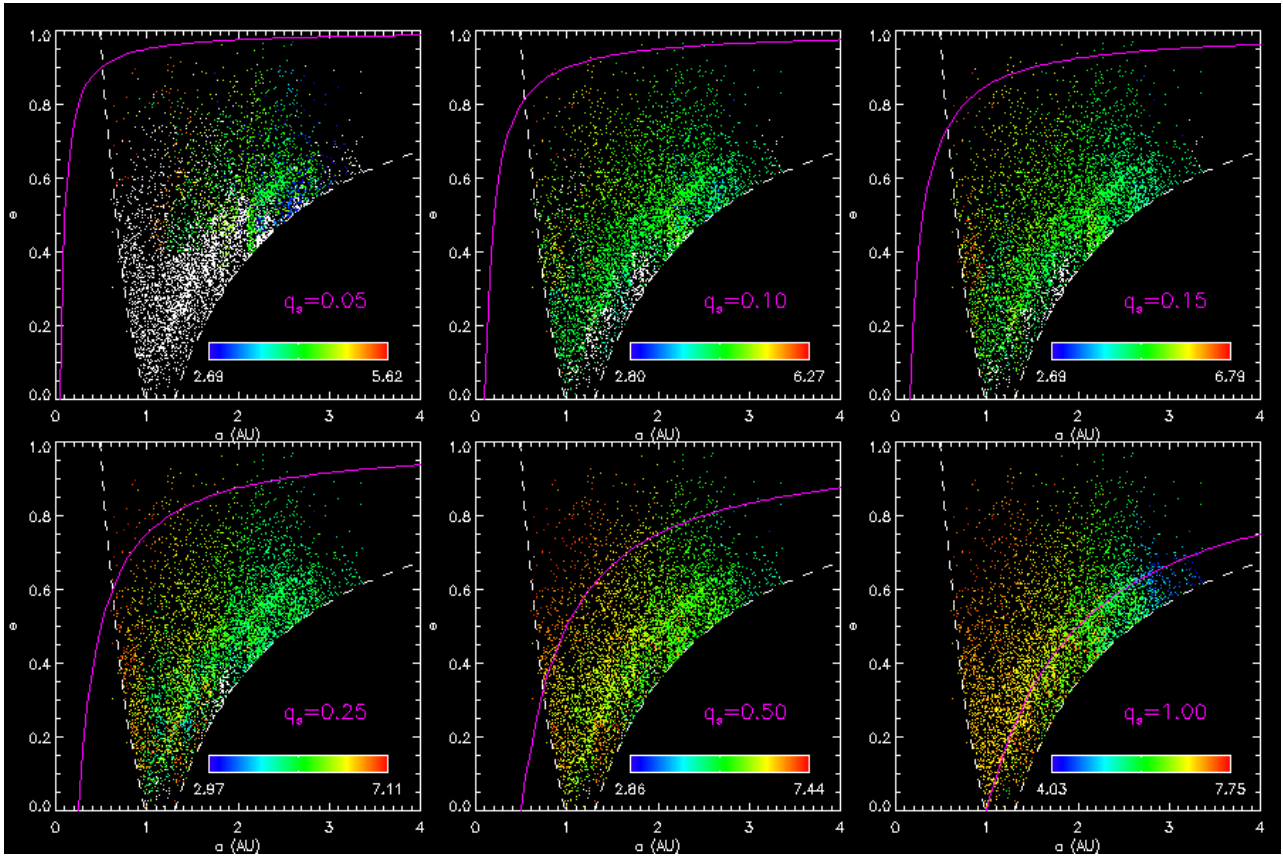}   
\vspace{0.1cm}
\caption{Same  as Fig.\ref{a_e_prob}, with  colors coded  according to
  the  logarithm of  cumulative  time spent  below  $q_s$ (white  dots
  indicates $T_s=0$).}
\label{a_e_temp}
\end{figure*}

\begin{figure*}
\includegraphics[width=10cm]{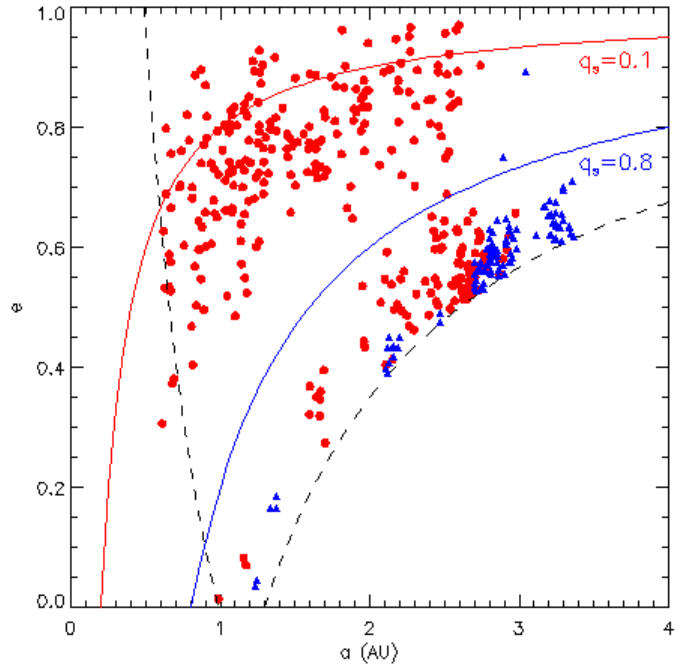}   
\vspace{0.1cm}
\caption{The disposition  of the ``hot''  and ``cool'' NEOs.   The hot
  NEOs (red circles)  are defined as those having  a probability of at
  least 50\% to  fall below 0.1~AU. The ``cool''  (blue triangles) are
  those having a probability of less than 10\% to fall below 0.8~AU.}
\label{hot_cool}
\end{figure*}

In Figure \ref{a_e_prob}, we report  the $a-e$ scatter plot of the NEO
population for several  values of $q_s$. For each  panel, the color of
NEOs is  coded according to  the probability of  experiencing $q<q_s$.
The curve $q=q_s$  is also overplotted. The figure  clearly shows that
the orbital  paths followed by each  NEO very often have  led to small
perihelion  distances,  much  shorter  than  their  present  $q$.   In
particular, this  is evident for $q_s=0.05,0.1$~AU: only  few NEOs are
presently below  those values, but  a significant fraction of  the NEO
population  spent some  time,  during the  evolution,  with a  smaller
perihelion.    The  average   probability   vs  $q_s$   is  shown   in
fig. \ref{prob_temp} (left panel).\\
It  is  obvious  that the  present  orbit  and  the past  history  are
correlated (this  is the rationale of our  computations); however, the
correlation is  not trivial.  As  a general trend,  $P(q_s)$ decreases
for objects with increasing value of their current perihelion distance
$q$.   Only bodies  with semi-major  axes in  the region  of  the main
resonances $\nu_6$ and 3:1 ($a=2-2.5$~AU) can have large $P(q_s)$ even
if   they  currently   have   a   large  value   of   $q$  (see   fig.
\ref{a_e_prob}).  This  happens because these  resonances entail large
oscillations of the eccentricity.  Moreover, we remark that the bodies
with the  smallest value of  $P(q_s)$ are those with  large semi-major
axis (beyond 2.5~AU),  mostly coming from the OB  source.  We may thus
expect that a significant percentage of NEOs, originating in the inner
part of  the Main  Belt and delivered  through the  $\nu_6$ resonance,
have been  affected by  the consequences of  the heating  process, and
thus  possibly physically  altered  (see also  the  discussion in  the
following).   The other  bodies should  have kept  more systematically
their original properties, at  least for what concerns the alterations
due to heating.
Another  interesting  result  concerns  the time  spent  at  distances
smaller  than  $q_s$.   In  Figure  \ref{a_e_temp}, we  show  the  NEO
population with colors coded  according to the cumulative time elapsed
for $q<q_s$.  We found that, even for small $q_s$, the time elapsed by
NEOs at close distances to  the Sun can be considerably high, reaching
the 10\% of  the typical life times (10~Myr) or, in  a few cases, even
more.   The average  cumulative time  spent  below $q_s$  is shown  in
fig. \ref{prob_temp} (right panel).\\
A closer look  at the results of our simulations  in order to identify
the most heated NEOs shows that 1\% of NEOs have a 50\% probability to
have  been at  $q<0.1$~AU.  Such  probability becomes  5\%  at 0.2~AU.
Their   cumulative  time spent  below  such   distances  can   vary
considerably, from  several $10^3$~yr to  several Myr.  Interestingly,
one of  the object  predicted to be  highly processed is  the peculiar
3200 Phaethon,  whose probability to  have been at $q<0.1$~AU  is 56\%
for a cumulative time of 0.3~Myr.\\
On the  other hand, there are NEOs  that never went close  to the Sun,
and therefore they represent the least processed bodies. For instance,
2\% of  NEOs have a probability of  less than 10\% to  have spent time
below $q_s=0.8$~AU.   These two  extreme sample of  NEOs -the  hot and
cool  bodies-  show  interesting  features  in  the  $a-e$  plot  (see
fig. \ref{hot_cool}).


\section{Thermal properties of NEOs}
\label{S:thermalproperties}

\begin{figure*}
\includegraphics[width=9cm,angle=90]{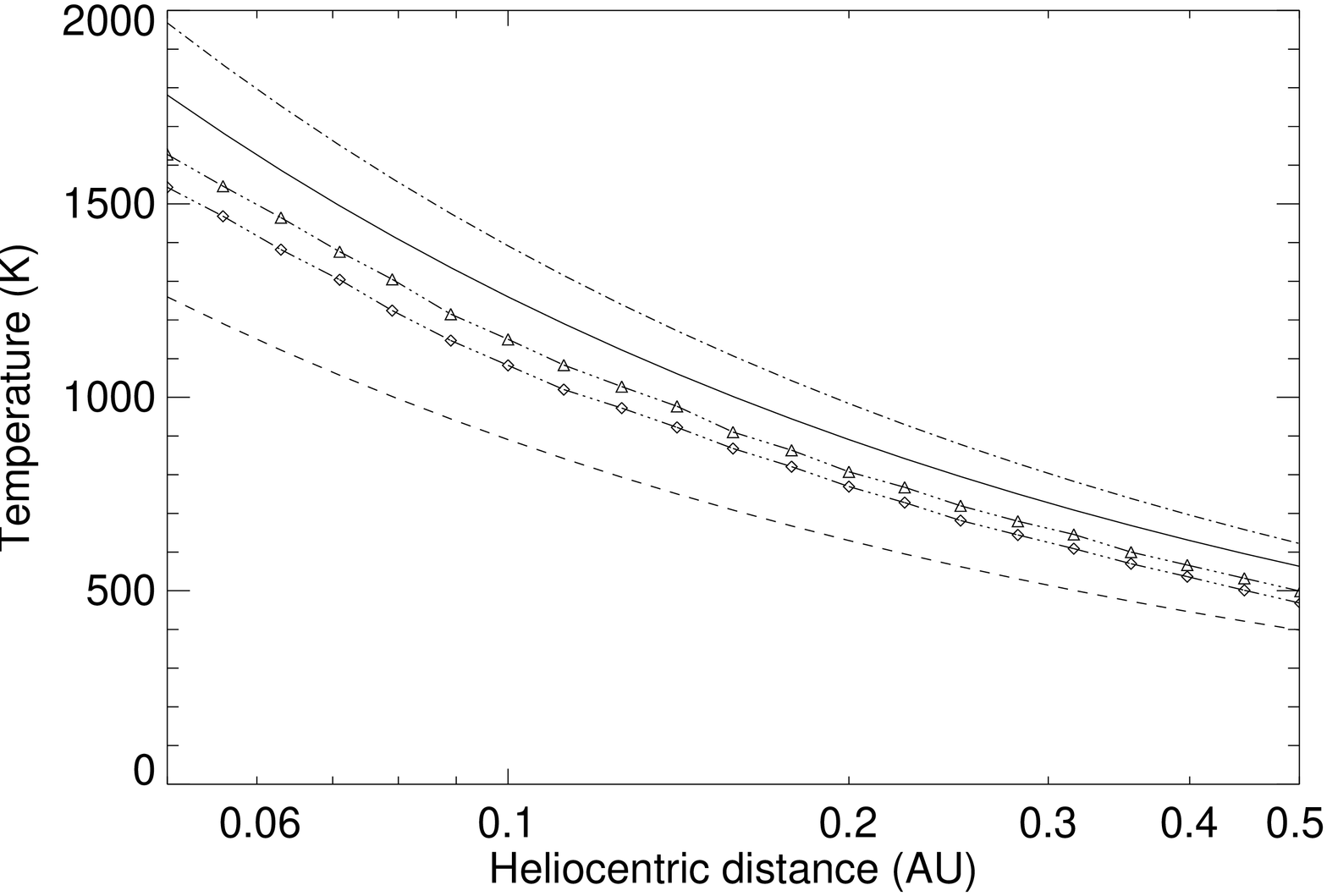}   
\vspace{0.1cm}
\caption{Temperature  of  NEOs   as  function  of  their  heliocentric
  distances. In particular: {\bf  continuous line}: temperature of the
  sub-solar  point  for  a   flat  surface  in  instantaneous  thermal
  equilibrium with sun light (physical parameters are $A$=0.05, $\eta$
  = 1.0 and $\epsilon=0.9$).  {\bf Dashed-dotted line}: temperature of
  the sub-solar point for a rough, low-albedo surface in instantaneous
  thermal  equilibrium   with  sun  light   (physical  parameters  are
  $A$=0.01,  $\eta$ =  0.6  and $\epsilon=0.9$).   {\bf Dashed  line}:
  temperature   of  an   isothermal   smooth-surface  body   (physical
  parameters are  $A$=0.05 $\eta$ = 4.0,  $\epsilon=0.9$).  This curve
  can represent the tempuerature of the meteorites.  {\bf Dotted lines
    with symbols}: temperature above  which a 70\% (diamonds) and 50\%
  (triangles) fraction of the surface area of an NEO is heated to. See
  section \ref{S:thermalproperties} for further details.}
\label{F:qtss}
\end{figure*}

In  the previous  section we  have shown  that a  fraction of  the NEO
population has  experienced small perihelion distances;  some of these
bodies had $q<q_s$ also for a long cumulative time.  For these objects
the  effects of  the  solar  heating may  have  altered their  surface
properties. In this section we use asteroid thermal and thermophysical
models to  investigate in  more detail the  dependence of  the surface
temperature of a body  on its heliocentric distance and thermophysical
properties.   Obviously  the presence  of  surface  regolith, and  its
continuous   mixing,   due   to   microcollisions  or   tidal   mixing
\citep{mar06a}  or the  YORP  reshaping \citep{har09},  may -at  least
partially- mask the above mentioned alterations.\\
The surface  temperature depends on the body's  thermal and rotational
properties.   However,  rough  estimate  of  surface  temperatures  at
perihelion can be obtained  from the equation of instantaneous thermal
equilibrium with sunlight:
\begin{equation}
T=\left[ (1-A)S_\odot q^{-2} \epsilon^{-1} \eta^{-1} \sigma^{-1} \right]^{0.25}
\label{eq:theq}
\end{equation}
where  $A$ is  the  bolometric  Bond albedo,  $S_\odot$  is the  solar
constant  at  1~AU  (1329  $Wm^{-2}$  ), $\epsilon$  is  the  infrared
emissivity,  $\sigma$  the Stefan-Boltzmann  constant  and $\eta$  the
so-called     beaming     parameter     \citep[see][and     references
  therein]{har02}.   The  latter can  be  seen  as  a measure  of  the
departure  of the  asteroid temperature  distribution from  that  of a
spherical,  smooth  body  with  all surface  points  in  instantaneous
thermal  equilibrium with  sunlight (which  would have  $\eta=1$). The
value of $\eta$ is a  strong function of the macroscopic roughness and
the  thermal inertia  of the  surface  \citep{spe89,har98,del07}.  The
latter  parameter, which  measures  the resistance  of  a material  to
temperature  changes,  is defined  as  $\Gamma=\sqrt{\rho \kappa  c}$,
where $\Gamma$  is the thermal inertia,  $\rho$ is the  density of the
material, $c$  is its specific  heat content and $\kappa$  the thermal
conductivity.  In  particular, $\eta >$  1 indicates a surface  with a
temperature  cooler   than  the  one  of   the  instantaneous  thermal
equilibrium, whereas  $\eta <$  1 is expected  for a surface  with low
thermal inertia and significant roughness.\\
An  upper limit for  the NEO  surface temperature  can be  obtained by
setting  in   Eq.   (\ref{eq:theq})  a  very   low  bolometric  albedo
($A$=0.01, corresponding to a geometric visible albedo, $p_V$, of 0.03
and the  default phase integral -G=0.15,  \cite{bow89}-; $p_V=0.03$ is
at  present the  lowest geometric  albedo measured  for NEAs),  a zero
thermal     inertia     and      a     strong     infrared     beaming
\citep[$\eta$=0.57,][]{wol05}.   For  an  infrared emissivity  of  0.9
-commonly     adopted      for     the     mineralogy      of     NEOs
\citep{lim05,sal91,mus97}- we find temperatures  of 1968 and 1391~K at
0.05~AU and  0.10~AU respectively.  Average parameter  values for NEOs
(i.e. $A$=0.05 from $p_V$=0.14 \citep{stu04} and G=0.15; $\eta$ = 1.0)
yields  subsolar  temperatures  of  1782  and 1260~K  at  0.05~AU  and
0.10~AU, respectively.  A lower  limit for the surface temperature can
be obtained  for a small  body with instantaneous  re--distribution of
the  heat  in  the  whole  volume,  with  no  difference  between  the
temperature  of day  and  night: namely  an  isothermal object.   This
latter case  can be obtained  using $\eta$=4 in  Eq.  (\ref{eq:theq}).
We obtain  1343~K at  0.05~AU and 949~K  at 0.10~AU for  $A$=0.05.  We
note  that the largest  ever measured  $\eta$-value for  a NEO  is 3.1
\citep[3671 Dionysus;][]{har99}.  Figure \ref{F:qtss} shows the surface
temperature as  function of the  heliocentric distances for  the three
cases described above.\\
However, observable consequences  of heating can come out  if and only
if the whole surface (or a  large amount of it) has been involved, and
thus altered.  In order to  estimate the fraction of the surface above
a  certain  temperature  threshold,  more accurate  knowledge  of  the
temperature  distribution  on  the  surface of  asteroids  is  needed.
Thermophysical  modeling is  required to  obtain this  information.  A
thermophysical      model     \citep[TPM,      see][and     references
  therein]{har02,spe89,lag96,eme98,del09}  describes an asteroid  as a
polyhedron made by  a mesh of planar facets.   The temperature of each
facet is  determined by  numerically solving the  one-dimensional heat
diffusion equation into the  subsurface with boundary conditions given
by the diurnal variable illumination and by the energy irradiated away
by each  facet at  the surface, and  by setting  the heat flow  at the
deepest subsurface element equal to zero. In our implementation of the
TPM, the subsurface  is divided into 32 slabs  of thickness 0.25 times
the  diurnal  heat  penetration depth  ($l_s$=$\sqrt{\kappa  \rho^{-1}
  c^{-1} \omega^{-1}}$), where $\omega = 2\pi/P$ with $P$ the rotation
period of the asteroid; the  deepest subsurface element is therefore 8
$l_s$ beneath  the surface.   Surface and subsurface  temperatures are
controlled  by   a  number   of  physical  parameters   including  the
heliocentric  distance  $q$,   $A$,  $\epsilon$,  $\Gamma$,  $P$,  the
direction of the rotation axis with  respect to the Sun, and the shape
of the body.\\
In the following, we assume  a spherical asteroid with the spin vector
perpendicular  to  its  orbital  plane, $P$=6  hours,  $\epsilon$=0.9,
$A$=0.1, $\Gamma$=200  $J m^{-2} s^{-0.5} K^{-1}$.  The  latter is the
average   value   of   the   thermal   inertia   for   km-sized   NEAs
\citep{del07}. The  TPM was run  on circular orbits  with heliocentric
distances ranging  from 0.05 AU  to 0.5 AU.  At the beginning  of each
run,  a complete rotation  of the  body is  performed and  the average
temperature  of  each  facet  is  recorded. The  temperature  at  each
rotation  step  (usually 360  steps  per  rotation  are performed)  is
calculated assuming instantaneous  thermal equilibrium between thermal
infrared emission and absorption  of solar energy.  (i.e.  $T_i=\left[
  (1-A)S_\odot  q^{-2}   \cos{\theta_i(t)}  \epsilon^{-1}  \sigma^{-1}
  \right]^{0.25}$),  where $T_i$  is the  temperature of  the $i_{th}$
surface element of  the mesh, $q$ is the  heliocentric distance of the
body, and $\theta_i$ is the angle formed by the normal of the $i_{th}$
surface element  of the  mesh with  the direction to  the Sun,  and it
varies with  the time $t$.  The  average temperature of  each facet is
then  used as  the  initial  temperature of  all  32 subsurface  slabs
including the one at the surface.
The  temperature profile  as a  function  of time,  i.e. the  asteroid
rotational phase, is monitored on each facet during the warm up phase,
which  can take  up to  50-100 (depending  on the  value of  $q$) full
rotations  until the  temperature profile  stabilizes and  the initial
temperature conditions are forgotten.
A final rotation is then performed and the maximum temperature of each
facet is stored. The areas  of those facets whose maximum temperatures
are found above a given temperature threshold ($T_s$) are added up and
the  ratio of  this area  to the  total area  of the  body  surface is
calculated for  a range  of values of  $T_s$ from  0 up to  2000K. The
value  of $T_s$  for  which the  ratio  of the  surface whose  maximum
temperature  is above  $T_s$  is equal  to  0.7 and  0.5  is found  by
interpolation.  Figure  \ref{F:qtss} shows  these values of  $T_s$ for
different   heliocentric  distances:   these   values  represent   the
temperature above which  a 70\% and 50\% fraction  of the surface area
of the body was heated to. \\
We also  note that the temperature  above were calculated  using a TPM
with  smooth  surface.  However,  it  is  well  known  that  roughness
increases the  average surface temperatures  by a maximum of  20-30 \%
with respect to those of a smooth surface.
Running the TPM on orbits with substantial eccentricity does not
affect significantly the final result.

\section{Difference between NEOs and meteorites}

Meteorites are collected on the Earth surface after having passed some
time in a  NEO-like orbit.  Their history may be  various: some may be
relatively young fragments  from cratering or catastrophic collisions,
or from tidal shattering of NEOs, others may have a longer independent
history,  maybe a  long orbital  evolution  as NEOs.   Despite of  the
different size, observed  NEOs and meteorites are believed  to share a
common  origin;  moreover  the  orbital  evolution  of  meteorites  is
typically dominated by  the same dynamical processes as  NEOs. \\
%
However, for what  concerns the alterations due to  heating, we expect
some important differences with  respect to NEOs: (i) Meteorites might
have spent  in near-Earth space a  shorter time than  the average NEO,
because of their short collisional  lifetime. This is suggested by the
fact that the statistics of falls  times (in the afternoon vs.  in the
morning) is skewed towards afternoon falls with respect to that of the
expected NEO impacts  \citep{mor98}. As a result the  time spent might
have  been not enough  for considerable  heating; (ii)  Meteorites are
systematically  by  far smaller  than  NEOs.   Thus  also the  thermal
history is likely different; (iii) bodies that collides with the Earth
(like meteorites) have  preferentially $q\sim1$~AU.  This introduces a
further bias with  respect to the observed NEO  population, where many
of the bodies have a very low probability of collision with the Earth.
Indeed,  since $P(q_s)$  decreases with  increasing $q$,  this implies
that meteorites are likely less heated than NEOs.\\
In  the  limit  of  a  very  small  body  we  can  imagine  an  almost
instantaneous  re--distribution of the  heat in  the whole  volume, no
day--night difference, a quasi-isothermal  object.  If we use the same
formulas  and basic  assumptions  to explore  this  extreme case,  the
factor 4 between the surface of a spherical body and its cross section
introduces a  factor $4^{-1/4}$  in the attainable  temperature, which
consequently decreases  by $\sim$30\% (see Eq.   \ref{eq:theq} and the
discussion in the previous Section).   Real meteorites may behave in a
more complex way, but somehow similar to the above scenario.  In fact,
an accurate  thermal model suggests that  a thin layer,  in the region
close to the subsolar point, may attain large temperatures, similar to
those  estimated for  asteroids. However,  we expect  that  those thin
layers are systematically erased  during the approach of the meteorite
to the Earth, mainly due  to atmospheric friction.  Thus the meteorite
``as we see  it'', i.e. after we recover it on  the ground, retains no
effect of the strong surface heating, and its thermal evolution can be
analyzed  in  terms of  the  ``isothermal''  approximation.  Thus  the
thermal history  of NEO and meteorites can  be considerably different.
We  must  take this  consideration  into  account  when attempting  to
compare  laboratory measured  quantities on  meteorites  with analogue
quantities derived from remote sensing of asteroids.\\

\section{Discussion: consequences of the heating}

The purpose of the present paper  is to point out the importance of the
heating history  of NEOs and meteoroids,  due to very  close passes to
the Sun.  Even  if the general ideas seem  rather robust, the physical
consequences have to  be explored in deeper detail.   In fact a number
of interesting phenomena can be caused by the mentioned heating.
Generally speaking  the heating can drive  volatile release, therefore
the surface  composition of NEOs  and, to a lesser  extend, meteoroids
become progressively depleted of volatiles. In the case of temperature
higher than 1200~K, silicate sublimation is expected.\\
For temperatures  below the melting points of  the different silicates
known to  be present on  asteroid surfaces, it  is worth to  note that
thermal heating can  cause annealing: this is in  general a transition
from an amorphous  to a crystalline structure. The  total time that an
object  spent with  $q<q_s$ is  extremely  important in  this kind  of
process.   This is  because  the rate,  $k$,  at which  a material  is
transformed from  amorphous to crystalline is  function of temperature
via an  equation of the form:  $k=a \exp(-E_a/bT)$, where  $a$ and $b$
are  constants   determined  from  experiments   \citep[see  e.g.][and
  references therein]{bru03}  and $E_a$  is the activation  energy for
the  transition. The  latter depends  on  the chemical  nature of  the
silicate. \\
Concerning the  most ``primitive'' NEOs,  i.e. those belonging  to the
C-types, more  severe effects are foreseen.  For  instance the aqueous
altered  carbonaceous chondrites are  strongly affected  by moderately
high  temperatures  \citep{hir96}.  We  note  that  the percentage  of
C--type  NEOs  with aqueous  alteration  features  (in particular  the
0.7~\mum\  feature)   is  rather  high  among  MBAs,   of  about  50\%
\citep{bus02, car03}.  This percentage drops below 10\% for NEOs: only
3 objects over more then  30 C--type NEOs having visible spectroscopic
measurements  show  subtle  absorption  features at  0.7~\mum.   These
results have  been also  found by \citet{vil05}  on the basis  of ECAS
photometry.  On the  other hand, the hydration is  a common feature in
the  meteorite collection,  in  agreement to  what our  considerations
predict.  However, the  data are too few and  probably not unambiguous
enough  to support  the  explanation  of the  difference  of the  NEOs
hydration properties  solely in terms  of heating.  For  instance, one
(2002~NX18) of  the C--type NEOs  with hydration features  is probably
coming from the Outer Belt, thus has presumably been poorly exposed to
heating. A similar  result holds also for 162173,  while the dynamical
history of 2002~DH2 is less  clear.  An alternative explanation can be
that NEOs sample peculiar source regions.  \\
Similar  considerations apply  also  to the  cometary objects,  namely
those  NEOs coming  from  the JFC  source  region (about  70 NEOs  are
estimated to  originate in this way).  The very close  passages to the
Sun and high temperature heating  should have produced a rapid release
of all volatiles, leading to the stage of depleted cometary nuclei.\\
Another interesting  aspect is related to the  space weathering, which
is  expected  to  be much  more  efficient  in  proximity to  the  Sun
\citep{mar06b,pao07}. Changing from 1~AU to 0.1~AU increases the space
weathering  rate by  two  order of  magnitude.   Similar
considerations  might also  apply to  other effects  depending  on the
solar radiation, such as Yarkovsky and YORP.\\
Moreover, we  find that the  thermal histories of NEOs  and meteorites
may be considerably different, therefore the above mentioned processes
may introduce systematic differences in their physical properties.
A future  analysis of these points  will be worthwhile;  in general we
remark that  the effects  of close  passages to the  Sun, in  terms of
heating,  but  also  as   concerns  other  processes,  should  not  be
neglected,  especially when  evaluating the  targets for  future space
missions to NEOs, such as  the European/Japanese Marco Polo mission or
the USA Osiris mission.\\
%
%
%
%
%
%
%
%


\section*{Acknowledgments}

We   thank  J.~Brucato   for   helpful  comments,   and  the   referee
A.~R.~Dobrovolskis  for carefully  reading the  paper.  PP  thanks ASI
funds.


\begin{thebibliography}{99}
%
\bibitem[\protect\citeauthoryear{Bottke et al.}{2000}]{bot00} Bottke, W.~F., Jedicke,
R., Morbidelli, A., Petit, J.-M., \& Gladman, B.\ 2000, Science, 288,
2190
%
\bibitem[\protect\citeauthoryear{Bottke et al.}{2002}]{bot02} Bottke, W.~F.,
Morbidelli, A., Jedicke, R., Petit, J.-M., Levison, H.~F., Michel, P.,
\& Metcalfe, T.~S.\ 2002, Icarus, 156, 399 
%
%
\bibitem[Bowell et al.(1989)]{bow89}  Bowell, E., Hapke, B., Domingue,
  D., Lumme, K., Peltoniemi, J., \& Harris, A.~W.\ 1989, Asteroids II,
  524
%
\bibitem[\protect\citeauthoryear{Brucato et
al.}{2003}]{bru03} Brucato J.~R., Baratta G.~A., Colangeli
L., Mennella V., Strazzulla G., 2003, asdu.conf.
%
\bibitem[\protect\citeauthoryear{Bus  \&  Binzel}{2002}]{bus02}  Bus,  S.~J.,  \&
Binzel, R.~P.\ 2002, Icarus, 158, 106
%
\bibitem[Carvano et al.(2003)]{car03} Carvano, J.~M., 
Moth{\'e}-Diniz, T., \& Lazzaro, D.\ 2003, Icarus, 161, 356 
%
\bibitem[Delbo' et~al. (2007)]{del07}
Delbo', M., Dell'Oro, A., Harris, A.~W., Mottola, S., Mueller, M., Sep 2007.
  Thermal inertia of near-earth asteroids and implications for the magnitude of
  the Yarkovsky effect. Icarus 190, 236.
%
\bibitem[Delbo' and Tanga (2009)]{del09}
Delbo', M., Tanga, P., Feb 2009. Thermal inertia of main belt asteroids smaller
  than 100km from IRAS data. Planetary and Space Science 57~(2), 259--265.
%
\bibitem[Emery et~al. (1998)]{eme98}
Emery, J.~P., Sprague, A.~L., Witteborn, F.~C., Colwell, J.~E., Kozlowski, R.
  W.~H., Wooden, D.~H., Nov 1998. Mercury: Thermal modeling and mid-infrared
  (5-12 $\mu m$) observations. Icarus 136, 104.
%
\bibitem[Grimm  \&  McSween(1993)]{gri93}  Grimm, R.~E.,  \&  McSween,
H.~Y.\ 1993, Science, 259, 653
%
\bibitem[Harris (1998)]{har98}
Harris, A.~W., Feb 1998. A thermal model for near-earth asteroids. Icarus 131,
  291.
%
\bibitem[Harris and Davies (1999)]{har99}
Harris, A.~W., Davies, J.~K., Dec 1999. Physical characteristics of near-earth
  asteroids from thermal infrared spectrophotometry. Icarus 142, 464.
%
\bibitem[Harris and Lagerros (2002)]{har02}
Harris, A.~W., Lagerros, J. S.~V., Jan 2002. Asteroids in the thermal infrared.
  Asteroids III, 205.
%
\bibitem[Harris et al.(2009)]{har09} Harris, A.~W., Fahnestock, E.~G.,
  \& Pravec, P.\ 2009, Icarus, 199, 310
%
\bibitem[Hiroi et al.(1993)]{hir93} Hiroi, T., Pieters, 
C.~M., Zolensky, M.~E., \& Lipschutz, M.~E.\ 1993, Science, 261, 1016
%
\bibitem[Hiroi et al.(1996)]{hir96} Hiroi, T., Zolensky,
M.~E.,  Pieters, C.~M.,  \&  Lipschutz, M.~E.\  1996, Meteoritics  and
Planetary Science, 31, 321
%
\bibitem[Keil(2000)]{kei00} Keil, K.\ 2000, PSS, 48, 887 
%
%
%
\bibitem[Lagerros (1996)]{lag96}
Lagerros, J. S.~V., Nov 1996. Thermal physics of asteroids. ii. polarization of
  the thermal microwave emission from asteroids. A{\&}A 315, 625.
%
\bibitem[Lim  et~al. (2005)]{lim05}  Lim,  L.~F., McConnochie,  T.~H.,
  Bell, J.~F.,  Hayward, T.~L., Feb 2005. Thermal  infrared (8-13 $\mu
  m$)  spectra  of 29  asteroids:  the  cornell mid-infrared  asteroid
  spectroscopy (midas) survey. Icarus 173, 385.
%
\bibitem[Marchi et al.(2006a)]{mar06a} Marchi, S., Magrin, S., 
Nesvorn{\'y}, D., Paolicchi, P., \& Lazzarin, M.\ 2006a, MNRAS, 368,
L39 
%
\bibitem[Marchi et al.(2006b)]{mar06b} Marchi, S., Paolicchi, 
P., Lazzarin, M., \& Magrin, S.\ 2006b, AJ, 131, 1138 
%
\bibitem[Morbidelli \& Gladman(1998)]{mor98} Morbidelli, A., \&
  Gladman, B.\ 1998, Meteoritics and Planetary Science, 33, 999 
%
\bibitem[Mustard and Hays (1997)]{mus97}
Mustard, J.~F., Hays, J.~E., Jan 1997. Effects of hyperfine particles on
  reflectance spectra from 0.3 to 25 $\mu m$. Icarus 125, 145.
%
%
%
%
\bibitem[Paolicchi  et al.(2007)]{pao07}  Paolicchi, P.,
  Marchi,  S., Nesvorn{\'y}, D.,  Magrin, S.,  \& Lazzarin,  M.\ 2007,
  A\&A, 464, 1139
%
\bibitem[Rivkin et al.(2002)]{riv02} Rivkin, A.~S., Howell, 
E.~S., Vilas, F., \& Lebofsky, L.~A.\ 2002, Asteroids III, 235 
%
\bibitem[Salisbury et~al. (1991)]{sal91}
Salisbury, J.~W., D'Aria, D.~M., Jarosewich, E., Aug 1991. Midinfrared
  (2.5-13.5 $\mu m$) reflectance spectra of powdered stony meteorites. Icarus
  92, 280.
%
\bibitem[Spencer et~al. (1989)]{spe89}
Spencer, J.~R., Lebofsky, L.~A., Sykes, M.~V., Apr 1989. Systematic biases in
  radiometric diameter determinations. Icarus 78, 337.
%
\bibitem[Stuart and Binzel (2004)]{stu04}
Stuart, J.~S., Binzel, R.~P., Aug 2004. Bias-corrected population, size
  distribution, and impact hazard for the near-earth objects. Icarus 170, 295.
%
\bibitem[Vilas and  Gaffey (1989)]{vil89} Vilas,  F. and M.  J. Gaffey
  M.  J.\ 1989.  Phyllosilicate  Absorption Features  in MainBelt  and
  OuterBelt Asteroid Reflectance Spectra. Science 246, 790.
%
\bibitem[Vilas (2005)]{vil05} Vilas, F.\ 2005, 36th Annual 
Lunar and Planetary Science Conference, 36, 2033 
%
\bibitem[Wolters et~al. (2005)]{wol05}
Wolters, S.~D., Green, S.~F., McBride, N., Davies, J.~K., May 2005. Optical and
  thermal infrared observations of six near-earth asteroids in 2002. Icarus
  175, 92.
%
\bibitem[Zavodny et al.(2008)]{zav08} Zavodny, M., Jedicke, 
R., Beshore, E.~C., Bernardi, F., \& Larson, S.\ 2008, Icarus, 198, 284 
%
\end{thebibliography}

\end{document}